\documentclass[12pt]{article}
\usepackage{hyperref}
\usepackage[top=2cm,bottom=3cm,left=2cm,right=2cm]{geometry}
\usepackage{graphicx}
\usepackage{amsmath}
\usepackage{amssymb} 
\usepackage{tikz-cd}
\usepackage{comment}
\usepackage{subcaption}
\usepackage{tikz-feynman}
\tikzfeynmanset{compat=1.1.0}
\usetikzlibrary{decorations.pathreplacing}

\numberwithin{equation}{section}
\numberwithin{figure}{section}

\newcommand{\nn}{\nonumber\\}
\newcommand{\tab}{\hspace{.6cm}}
\newcommand{\tr}{\text{Tr}}
\newcommand{\tv}{\Psi_\text{tv}}
\newcommand{\no}[1]{:\, #1 \,:}

\begin{document}
	
	\begin{titlepage}

		\hfill \today
		\begin{center}
			\vskip 2cm
			
			{\Large \bf {Hamiltonian of open string field theory\\ for time-dependent solutions}
			}
			
			\vskip 0.5cm
			
			\vskip 1.0cm
			{\large {Vin\'{\i}cius Bernardes and Riccardo Poletti }}
			
			\vskip 0.5cm
			
			{\em  \hskip -.1truecm
				
				CEICO, FZU - Institute of Physics of the Czech Academy of Sciences \\
				Na Slovance 2, 182 21, Prague 8, Czech Republic\\
                
                \vskip 0.3cm
                
                Institute of Theoretical Physics, Faculty of Mathematics and Physics,
                Charles University, Prague, V Holešovičkách 2, 180 00 Prague 8, Czech Republic

				\vskip 0.5cm
				\tt \href{mailto:viniciusbernsilva@gmail.com}{viniciusbernsilva@gmail.com}, \href{mailto:polettiricc@gmail.com}{polettiricc@gmail.com} \vskip 5pt }
			
			\vskip 2.0cm
			{\bf Abstract}
			
		\end{center}
		\vskip 0.25cm
		\noindent
		\begin{narrower}
			\baselineskip15pt 
            
            Using the recently introduced expression for conserved charges of $L_\infty$ theories, we give an explicit form for the Hamiltonian of Witten's bosonic open string field theory. We evaluate it on the Erler-Maccaferri analytic solution, describing any D$p$-brane configuration associated to a BCFT. We use the expression to compute the energy of the rolling tachyon solution and find exact agreement with the boundary state result by Sen.
			
		\end{narrower}
	\end{titlepage}
	
	\tableofcontents
	\baselineskip15pt
	
\section{Introduction}

Witten's open bosonic string field theory describes the dynamics of open bosonic string fields, which are fluctuations of D-brane configurations around a reference background \cite{SenZwiebach:2024Review,Witten:OSFT}. Many solutions of the theory have been found and studied, including both perturbative and exact ones \cite{Erler:AnalyticLectures}. The Erler-Maccaferri solution \cite{erler-maccaferri,erler-maccaferri-2} provides the most general framework for constructing exact solutions based on the BCFT description of the open string backgrounds.

For static solutions, the energy evaluates to minus the on-shell Lagrangian. But for time-dependent solutions, this is not the case. An important example is the rolling tachyon solution, whose energy has been computed from the associated boundary state by Sen \cite{Sen:2002RollingTachyon}. The energy has been studied in SFT and related toy models \cite{Kudrna,Yang,Moeller:2002vx,Pressure}. The evaluation of the energy from the string field theory solution has been carried out in \cite{CMY}, using a perturbative solution in Siegel gauge for the rolling tachyon on a ZZ brane of the $c=1$ theory, and it matches with the boundary state result perturbatively. The computation uses the idea of a symplectic form for OSFT inspired by Witten \cite{Witten:SymplecticForm}, and it agrees with the Ellwood invariant prescription.

The symplectic form of OSFT was further developed in a series of recent works \cite{Bernardes:CovariantPhaseSpace,Bernardes:2026RollingTachyons,Bernardes:Lump,ElectricField,StettingerAli}, where Witten's proposal for the symplectic form is generalized to any interacting gauge theory described in terms of $L_\infty$ algebras. The new proposal was checked against a number of OSFT examples, including the perturbative rolling tachyon solution in Siegel gauge. Following these developments, an expression for the conserved charges was constructed in \cite{Bernardes:ConservedCharges}. See also \cite{KMPV} for a construction of surface charges based on the formalism of boundary terms in SFT \cite{MPRV,MRV,StettingerBoundary,Firat-Mamade}. The goal of this paper is to use the formalism of \cite{Bernardes:ConservedCharges} to write the Hamiltonian of the Witten OSFT and evaluate it on the Erler-Maccaferri solution for the rolling tachyon.

In Section \ref{sec:Witten} we review Witten's bosonic open string field theory, and the Erler-Maccaferri analytic solution, which describes open string backgrounds associated to any BCFT. In Section \ref{sec:hamiltonian} we review the conserved charges formalism. We use it to write down the Hamiltonian of Witten OSFT around a possibly time-dependent D$p$-brane configuration, which takes the form
\begin{align}
    H = & ~ \frac{1}{2} \tr_{Q\sigma} (\Psi i p_0 \Psi) + \frac{1}{2\pi i} \tr_{Q\sigma} ( \Psi (c\partial X^0 - \bar c \bar \partial X^0) ) + \frac{1}{6} \tr_{i p_0 \sigma} (\Psi Q \Psi),
    \label{eq:hamiltonian-witten-intro}
\end{align}
where $\Psi$ is the open string field, $Q$ is the BRST operator, $ip_0$ is the generator of time translations, and $\sigma$ is the sigmoid operator which effectively localizes the expression to a time slice. The operator $c\partial X^0 - \bar c \bar \partial X^0$ is a boundary insertion on the identity string field, and it does not vanish for an arbitrary time-dependent BCFT.

We evaluate the Hamiltonian on the Erler-Maccaferri solution associated to an arbitrary target BCFT, and find
\begin{align}
    H = & ~ - \frac{1}{2} \tr_{Q\sigma}^* ( \tv ip_0 \tv ) - \frac{1}{2\pi i} \tr_{Q\sigma}^* ( \tv (c\partial X^0 - \bar c \bar \partial X^0) ) - \frac{1}{6} \tr^*_{ip_0\sigma} (\tv Q \tv) \nn
    & ~ + \frac{1}{2} \tr_{Q\sigma} ( \tv ip_0 \tv ) + \frac{1}{2\pi i} \tr_{Q\sigma} ( \tv (c\partial X^0 - \bar c \bar \partial X^0) ) + \frac{1}{6} \tr_{ip_0\sigma} (\tv Q \tv),
    \label{eq:hamiltonian-introduction}
\end{align}
where $\tv$ is the Erler-Schnabl simple tachyon vacuum \cite{erler-schnabl-simple}. The trace $\tr(\cdots)$ is a correlator on the reference BCFT while $\tr^*(\cdots)$ is a correlator in the target BCFT. The construction requires a notion of Cauchy slice with respect to lightcone time $x^+ = x^0 + x^1$, so the expression holds under the assumption that the solution does not deform the $X^1$ sector. This expression does not depend on the boundary condition changing operators used to define the solutions. This avoids potential problems coming from singular marginal deformations, such as the rolling tachyon solution, for which the Erler-Maccaferri solution requires dealing with flag states.

In Section \ref{sec:rolling} we review the definition of the rolling tachyon BCFT due to Recknagel and Schomerus \cite{Recknagel-Schomerus}, which is formulated for a spacetime with Euclidean signature. However, the formalism of conserved charges requires the notion of a time slice along a non-compact direction. We assume that the Recknagel-Schomerus construction can be analytically continued to a Lorentzian spacetime, and compute the rolling tachyon energy by evaluating eq. \eqref{eq:hamiltonian-introduction} on it. We find the energy shift
\begin{align}
    \Delta E = & ~ - V_p T_p \sin^2 (\pi \lambda),
\end{align}
where $V_p$ is the D$p$-brane volume and $T_p$ is the D$p$-brane tension. This gives the total energy
\begin{align}
    E = & ~ V_p T_p + \Delta E = V_p T_p \cos^2 (\pi\lambda),
\end{align}
in agreement with Sen's boundary state result in \cite{Sen:2002RollingTachyon}.

\paragraph{Conventions:} We take $\alpha'=1$ and $g_o = 1$. We assume mostly plus Lorentzian signature for spacetime. The correlator of three $c$ ghosts on the upper half plane (UHP) is
\begin{align}
    \langle c(z_1) c(z_2) c(z_3) \rangle_\text{UHP} = (z_1-z_2)(z_1-z_3)(z_2-z_3).
\end{align}
The divergent D$p$-brane spacetime volume is $V_{p+1} = (2\pi)^{p+1} \delta^{p+1} (0)$ and the D$p$-brane tension is $T_p = 1/(2\pi^2)$.

\section{Witten OSFT and the Erler-Maccaferri solution}
\label{sec:Witten}

\subsection{Witten's open bosonic string field theory}

Witten's open bosonic string field theory is a field theory whose fields are elements of the Hilbert space of a bosonic open string worldsheet model. The theory can be constructed around different D$p$-brane configurations characterized by a BCFT which we refer to as the \emph{reference BCFT}. The dynamical fields are elements $\Psi$ of ghost number 1. The action can be written as
\begin{align}
    S = & ~ - \text{Tr} \bigg( \frac{1}{2} \Psi * Q \Psi + \frac{1}{3} \Psi * \Psi * \Psi \bigg),
    \label{eq:witten-action}
\end{align}
where the kinetic operator $Q$ is the worldsheet BRST operator, the star product $*$ maps a pair of string fields to another one, and is defined in terms of the geometry of the Witten vertex \cite{Witten:OSFT}, and the trace $\tr(\cdot * \cdot)$ is defined using the Witten integral, which can be related to the BPZ inner product via the Witten vertex.

The equation of motion can be written as $q_\Psi = 0$, where\footnote{We will generally omit the star symbol in the star product, writing $A*B=AB$.}
\begin{align}
    q_\Psi = Q \Psi + \Psi * \Psi.
\end{align}
The string field $\Psi$ can be written as a fluctuation $\psi$ around a solution $\Psi_*$, as $\Psi = \psi + \Psi_*$. The fluctuation satisfies the equation $Q_{\Psi_*} \psi + \psi*\psi = 0$, given in terms of the operator
\begin{align}
    Q_\Psi = Q + [\Psi, \cdot].
\end{align}
This is the non-linear generalization of the kinetic operator obtained by considering the interactions, which we refer to as the wave operator.

It is clear that Witten OSFT constructed around a time-independent BCFT is invariant under time translations. The translation generator is an operator that can be written as the contour integral of translation currents,
\begin{align}
    i p_0 = & ~ \int_C \frac{dz}{2\pi i} \partial X^0 (z) - \int_C \frac{d\bar z}{2\pi i} \bar \partial X^0 (\bar z).
    \label{eq:p0}
\end{align}
The integral over the anti-holomorphic part has a minus sign, which comes from the opposite orientation of the contour. The action is invariant under an infinitesimal transformation given by the Lie derivative
\begin{align}
    \mathcal{L}_t \Psi = & ~ i p_0 \Psi .
\end{align}
For a time-independent reference BCFT, $ip_0$ commutes with $Q$ and acts as a derivation of the star product, so the variation of the action is a total $ip_0$ derivative,
\begin{align}
    \mathcal{L}_t S = & ~ - \tr \bigg( ip_0 \Big( \frac{1}{2} \Psi Q \Psi + \frac{1}{3} \Psi^3 \Big) \bigg) = 0 ,
    \label{eq:time-translation}
\end{align}
which vanishes because the contours of $ip_0$ can be joined into a closed contour in the bulk.

If the reference BCFT is time-dependent, meaning that it describes a non-static D$p$-brane configuration, the time invariance of the theory is spontaneously broken. The theory is still invariant under the transformation
\begin{align}
    \mathcal{L}_t \Psi = & ~ ip_0 \Psi + \frac{1}{2\pi i} ( c\partial X^0 - \bar c \bar \partial X^0 ).
\end{align}
where $c \partial X^0 - \bar c \bar \partial X^0$ is understood as an insertion on the identity string field. We do not use the doubling trick because it depends on the choice of reference BCFT. The action transforms as
\begin{align}
    \mathcal{L}_t S = & ~ -\frac{1}{2} \tr( \Psi [Q,ip_0] \Psi ) -  \tr \bigg( \frac{1}{2\pi i} (c \partial X^0 - \bar c \bar \partial X^0) \Psi^2 \bigg).
\end{align}
Notice that the operator $[Q,ip_0]$ acts as the contour integral
\begin{align}
    [Q,ip_0] = & ~ \int \frac{dz}{2\pi i} \partial ( c \partial X^0 ) - \int \frac{d\bar z}{2\pi i} \bar \partial ( \bar c \bar \partial X^0 ).
\end{align}
Again, in an arbitrary BCFT we cannot use the doubling trick, so the integrals run over the same contour with endpoints on the boundary. Since the integrand is a total derivative, we can evaluate it on the endpoints, and the variation of the action becomes
\begin{align}
    \mathcal{L}_t S = & ~ \frac{1}{2} \tr \bigg( \Psi \frac{1}{2\pi i} (c\partial X^0 - \bar c \bar \partial X^0) \Psi \bigg) + \frac{1}{2} \tr \bigg( \Psi^2 \frac{1}{2\pi i} (c\partial X^0 - \bar c \bar \partial X^0)  \bigg) \nn
    & ~ - \tr \bigg( \frac{1}{2\pi i} (c \partial X^0 - \bar c \bar \partial X^0) \Psi^2 \bigg) = 0.
\end{align}
We can see that all terms cancel using cyclicity.

The conserved charge associated with this symmetry is the Hamiltonian of the theory. We explain how to construct it in Section \ref{sec:hamiltonian}.

\subsection{Erler-Maccaferri solution}

The Erler-Maccaferri solution \cite{erler-maccaferri, erler-maccaferri-2} for a given open string background is expressed in terms of its associated pair of boundary condition changing operators. It is constructed in terms of the tachyon vacuum solution \cite{SchnablTachyonVacuum,erler-schnabl-simple} and a pair of \emph{intertwining fields} $\Sigma$, $\bar \Sigma$, and it is given by
\begin{align}
    \Psi = & ~ \Psi_\text{tv} - \Sigma \Psi_\text{tv} \bar \Sigma.
    \label{eq:em-solution}
\end{align}
The simple tachyon vacuum solution takes the form\footnote{We make the choice, for later convenience, of putting the ``security strip'' $1/(1+K)$ on the right end of $\tv$, instead of placing $1/\sqrt{1+K}$ to each side.}
\begin{align}
    \Psi_\text{tv} = & ~ c(1+K) Bc \frac{1}{1+K}.
    \label{eq:tachyon-vacuum}
\end{align}
It is expressed in terms of the string fields $K$, $B$ and $c$ \cite{Erler:AnalyticLectures}. The field $K$ has ghost number zero and even parity, while the fields $B$ and $c$ have odd parity and ghost numbers $-1$ and $1$, respectively. Some key properties that they satisfy are
\begin{align}
    [K,B] = & ~ 0, \tab [B,c] = 1, \tab [K,c] = \partial c, \nn
    B^2 = & ~ 0, \tab c^2 = 0, \nn
    QB = & ~ K, \tab QK = 0, \tab Qc = cKc,
\end{align}
where $[\cdot,\cdot]$ are graded commutators. A wedge state of width $\alpha$ is a string field generated by $K$, which can be expressed as $\Omega^\alpha = e^{-\alpha K}$. On the sliver frame it is represented as a patch $z=x+iy$ for $0<x<\alpha$ and $0\leq y < \infty$. The boundary is located at $y=0$. Under the star product, the wedge states add their width, $\Omega^\alpha \Omega^\beta = \Omega^{\alpha+\beta}$. The wedge state of width zero is the identity string field. The fields $K$, $B$ and $c$ can be expressed as the identity string field with associated operator insertions. Since $\Psi_\text{tv}$ is a solution, it satisfies
\begin{align}
    Q\Psi_\text{tv} + \Psi_\text{tv}^2 = 0.
\end{align}

The intertwining fields $\Sigma, \bar \Sigma$ each contain an insertion of a boundary condition changing operator. The solution $\Psi$ is an element of the Hilbert space of the \emph{reference BCFT}, describing the original D$p$-brane configuration. We can see that $\tv$ appears in two different places in eq. \eqref{eq:em-solution}: in the first term it appears by itself, and it is understood to be an element of the Hilbert space of the reference BCFT; but in the second term it appears in between the intertwining fields, so it should be understood as an element of the Hilbert space of the \emph{target BCFT}. When manipulating expressions involving the Erler-Maccaferri solution, we must keep track of which BCFT Hilbert space the string fields belong to. If the expression involves correlators, we must specify in which BCFT these correlators are evaluated.

\begin{figure}[t]
    \centering
    \includegraphics[width=\textwidth]{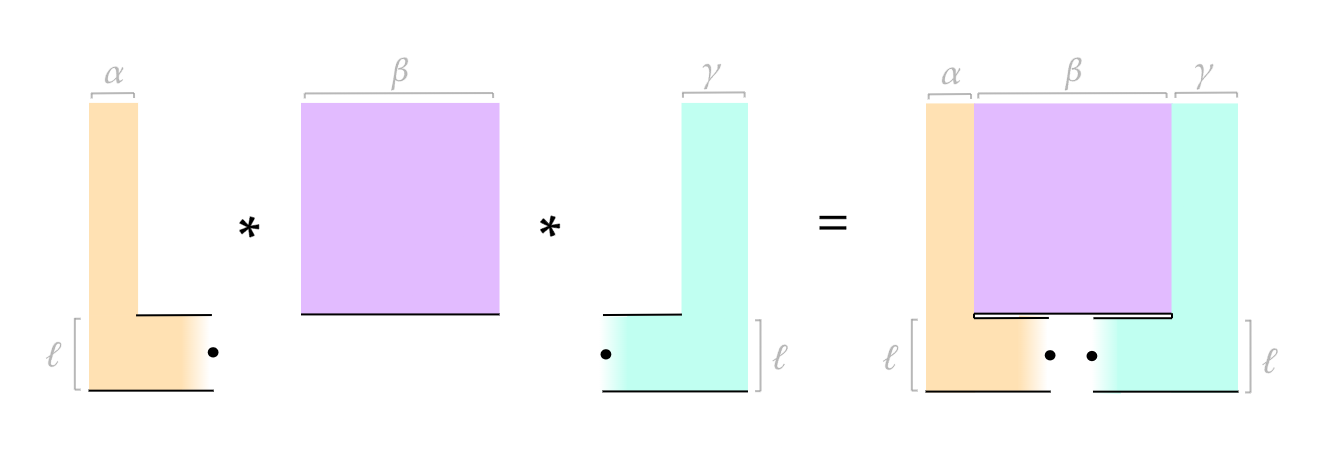}
    \caption{Multiplication of a flag state of width $\alpha$, a wedge state of width $\beta$ and an anti-flag state of width $\gamma$ in the sliver frame. The boundaries are represented as black lines around the colored surfaces. Insertions at the tip of the flag are represented by black dots.}
    \label{fig:flag-states}
\end{figure}

If the BCFT associated to the solution is a regular marginal deformation of the reference BCFT, meaning that it has a regular self-OPE, the intertwining fields are a sum of wedge states with particular operator insertions. On the other hand, if the target BCFT is singular, the intertwining fields are described by a different kind of surface states, the flag states \cite{erler-maccaferri-2}, which regularize the OPEs of boundary condition changing operators on the solution. The flag state can be seen as a wedge state with an additional patch that extends the boundary to $z=-\infty$. The extended part of the surface is the ``flag'', whereas the wedge part is the ``pole''. The flag has a width $\ell > 0$. Following the left-handed convention \cite{Erler:AnalyticLectures} (i.e. the real axis is taken to point left instead of right), we can describe the patch with the coordinates
\begin{align}
    z = & ~ x + i y, \tab 0<x<\alpha, \tab\tab 0 \leq y < \infty, \tab~ \text{Im} z \geq 0, \nn
    z = & ~ x + i y, \tab -\infty < x < \alpha, \tab - \ell < y < 0, \tab~ \text{Im} z < 0.
\end{align}
The anti-flag state is defined analogously, with the flag extending to $z=\infty$,
\begin{align}
    z = & ~ x + i y, \tab 0<x<\alpha, \tab~ 0 \leq y < \infty, \tab\, \text{Im} z \geq 0, \nn
    z = & ~ x + i y, \tab 0 < x < \infty, \tab - \ell < y < 0, \tab \text{Im} z < 0.
\end{align}
We denote by $| \mathcal{O} )$ the flag state with vanishing wedge width, $\alpha=0$, and with the insertion of the operator $\mathcal{O}$ at the point $z=-\infty-i \ell$ (the tip of the flag). Analogously, $( \mathcal{O}|$ is the anti-flag state with vanishing wedge width and the insertion of $\mathcal{O}$ at the point $z=\infty-i\ell$. Figure \ref{fig:flag-states} depicts the multiplication of a flag state, a wedge state and an anti-flag state in the sliver frame.

The intertwining fields are defined as
\begin{align}
	\Sigma = & ~ Q_{\Psi_\text{tv}} \bigg( B |\sigma) \frac{1}{1+K} \bigg), \tab
	\bar \Sigma = Q_{\Psi_\text{tv}} \bigg( (\bar \sigma| \frac{B}{1+K} \bigg).
    \label{eq:intertwining-fields}
\end{align}
Here $\sigma$ and $\bar \sigma$ mean the boundary condition changing operators, which are inserted at the tip of the flags. We write them explicitly only in this subsection, to avoid confusion with the sigmoid operator. One can check that the intertwining fields satisfy
\begin{align}
    \bar \Sigma \Sigma = 1,
\end{align}
using $Q_{\tv}^2=0$ and $(\bar \sigma | \sigma ) = 1$. The product $\Sigma \bar \Sigma $, in the reverse order, is a projector: $(\Sigma \bar \Sigma) (\Sigma \bar \Sigma) = \Sigma \bar \Sigma$. Because $\Sigma$ and $\bar \Sigma$ are $Q_{\tv}$-closed, they satisfy
\begin{align}
    Q \Sigma = & ~ - \Psi_\text{tv} \Sigma + \Sigma \Psi_\text{tv}, \tab
    Q \bar \Sigma = - \Psi_\text{tv} \bar \Sigma + \bar \Sigma \Psi_\text{tv}.
    \label{eq:intertwining-brst}
\end{align}
These properties are enough to show that \eqref{eq:em-solution} solves the equation of motion.

A subtlety of the flag state geometry requires that the $c$ ghost insertions of $\tv$ in eq. \eqref{eq:intertwining-fields} are placed at a finite distance above the boundary \cite{erler-maccaferri-2}. When the pairs of flag and anti-flag states cancel inside a correlator, and the resulting geometry is simply a cylinder, we can place the $c$ ghosts at the boundary.

\section{The Hamiltonian}
\label{sec:hamiltonian}

\subsection{Conserved charges in OSFT}

The expression for the conserved charge associated to a given symmetry of any gauge theory with the structure of an $L_\infty$ algebra was derived in \cite{Bernardes:ConservedCharges}. We will review it briefly and use it to construct the Hamiltonian.

Let $\Psi$ denote the field of the theory, $q_\Psi$ the equations of motion, and $Q_\Psi$ the associated wave operator. We denote the inner product of the theory as a trace. A symmetry of the theory can be described by a \emph{generating state} $\xi_\Psi$, which is a function of $\Psi$ describing the symmetry as $\Psi' = \Psi + \xi_\Psi$. The symmetry can also be described by a \emph{generating operator} $\Xi_\Psi$, which is related to the generating state by $\delta \xi_\Psi = \Xi_\Psi \delta \Psi$, where $\delta$ is a differential in the space of fields. The conserved charge associated to the symmetry is defined by
\begin{align}
    F_\xi = & ~ - \int_0^1 ds \tr \bigg( \frac{\partial \Psi (s)}{\partial s} \Big( [Q_{\Psi(s)}, \sigma] \xi_{\Psi(s)} - [\Xi_{\Psi(s)}, \sigma] q_{\Psi(s)} \Big) \bigg).
    \label{eq:conserved-charge}
\end{align}
The function $\Psi(s)$ is an interpolation between $s=0$ and $s=1$, where $\Psi(0)=0$, and $\Psi(1)=\Psi$ is the physical field. The integrand in eq. \eqref{eq:conserved-charge} is always a total $s$ derivative, so the expression depends only on the physical field. Notice that we cannot use the equations of motion for $\Psi(s)$, so $q_{\Psi(s)} \neq 0$.

The \emph{sigmoid operator} $\sigma$ can be seen as a function of time and interpolates between 0 and 1,
\begin{align}
    \lim_{t\to -\infty} \sigma(t) = 0, \tab \lim_{t\to \infty} \sigma(t) = 1.
    \label{eq:sigmoid-condition}
\end{align}
The conserved charge is independent of the particular choice of sigmoid, as long as it satisfies this asymptotic behavior. And it defines a notion of a fuzzy time slice, so the independence of eq. \eqref{eq:conserved-charge} of the choice of sigmoid can be seen as the statement that the charge is conserved in time.

For Witten OSFT, we can take the sigmoid to be an operator that acts as a multiplication by a function of a zero mode of the free scalar boson. A technically convenient choice of sigmoid for analytic manipulations of Witten OSFT is the \emph{midpoint lightcone sigmoid} \cite{Bernardes:Lump}. We take the time variable to be a zero mode in the expansion around the midpoint, so its relation with the Witten product takes a simple form. We write the operator as
\begin{align}
    \sigma = & ~ \int_{-\infty}^\infty \frac{dE}{2\pi} \sigma(E) e^{i E X^+ (i\infty, \bar i \infty)}
    \label{eq:midpoint-sigmoid}
\end{align}
in the sliver frame. The function $\sigma(E)$ is the Fourier transform of any function of time for which the sigmoid satisfies the asymptotic conditions \eqref{eq:sigmoid-condition}. In the momentum space, this condition is $\dot \sigma(E=0) = 1$. We use the lightcone time coordinate $X^+ = X^0 + X^1$, so the vertex operator associated with the sigmoid has zero conformal weight---the Lorentzian signature provides different signs for the conformal weight coming from the $X^0$ and $X^1$ sectors, so they cancel.

Now we can use eq. \eqref{eq:conserved-charge} to write the Hamiltonian. We consider a possibly time-dependent reference BCFT, so time translation is described by the generating state and operator
\begin{align}
    \xi_\Psi = & ~ ip_0 \Psi + \frac{1}{2\pi i} (c\partial X^0 - \bar c \bar \partial X^0), \tab 
    \Xi_\Psi = i p_0.
\end{align}
The Hamiltonian takes the form
\begin{align}
    H = & ~ - \int_0^1 ds \tr \bigg( \frac{\partial \Psi (s)}{\partial s} \Big( [Q_{\Psi(s)}, \sigma] ip_0 \Psi(s) + \frac{1}{2\pi i} [Q_{\Psi(s)},\sigma] (c\partial X^0 - \bar c \bar \partial X^0) - [ip_0,\sigma] q_{\Psi(s)}  \Big) \bigg).
    \label{eq:hamiltonian-s-integral}
\end{align}
The choice of midpoint sigmoid is special because it commutes with the star product. One way to see that is by looking at its representation in the basis of wedge states. The sigmoid operator can be thought of as the multiplication by the identity string field with the associated midpoint insertion. The multiplication is the same whether done by the left or by the right,
\begin{align}
    \sigma * \Psi = \Psi * \sigma.
\end{align}
So in eq. \eqref{eq:hamiltonian-s-integral} we can use the relation
\begin{align}
    [Q_{\Psi},\sigma] = & ~ [Q,\sigma] + [[\Psi,\cdot],\sigma] \nn
    = & ~ [Q,\sigma] + [\Psi, \cdot] \sigma - \sigma [\Psi, \cdot] \nn
    = & ~ [Q,\sigma].
\end{align} 
The operators $[Q,\sigma]$ and $[ip_0,\sigma]$ act as the midpoint insertion of $Q\sigma$ and $ip_0\sigma$, respectively. We denote a midpoint insertion as a subscript on ``Tr'',
\begin{align}
    H = & ~ \int_0^1 ds \bigg( \tr_{Q\sigma} \bigg( \frac{\partial \Psi(s)}{\partial s} ip_0 \Psi (s) \bigg) + \frac{1}{2\pi i} \tr_{Q\sigma} \bigg( \frac{\partial \Psi(s)}{\partial s} (c\partial X^0-\bar c \bar \partial X^0) \bigg) \nn
    & ~ \hspace{2cm} + \tr_{ip_0\sigma} \bigg( \frac{\partial \Psi(s)}{\partial s} Q \Psi (s) \bigg) + \tr_{ip_0\sigma} \bigg( \frac{\partial \Psi(s)}{\partial s} \Psi^2(s) \bigg) \bigg).
    \label{eq:hamiltonian-midpoint}
\end{align}
Acting with $ip_0$ on eq. \eqref{eq:midpoint-sigmoid} we find 
\begin{align}
    ip_0 \sigma = & ~ \int_{-\infty}^\infty \frac{dE}{2\pi} \sigma(E) ip_0 e^{iEX^+ (i\infty,\bar i\infty)} \nn
    = & ~ \int_{-\infty}^\infty \frac{dE}{2\pi} \underbrace{\sigma(E) i E}_{=\dot \sigma(E)}  e^{iEX^+ (i\infty,\bar i\infty)} .
\end{align}
The function $\dot \sigma(E)$ is the Fourier transform of the derivative of the sigmoid with respect to the time variable. The independence of the conserved charges of the choice of sigmoid implies that the only contribution comes from $E=0$, where $\dot \sigma(E=0) = 1$. Analogously, the midpoint insertion of $Q\sigma$ is
\begin{align}
    Q \sigma = & ~ \int_{-\infty}^\infty \frac{dE}{2\pi} \sigma(E) Q e^{iE X^+ (i\infty,\bar i\infty)} \nn
    = & ~ \int_{-\infty}^\infty \frac{dE}{2\pi} \dot \sigma(E) \no{ \Big( c(i\infty) \partial X^+ (i\infty) + \bar c(\bar i\infty) \bar \partial X^+ (\bar i\infty) \Big) e^{iE X^+ (i\infty, \bar i\infty) } }.
\end{align}
Here we can see why it is convenient to let the sigmoid depend on the lightcone time: the conformal weight of the plane wave is zero, so the term proportional to $\partial c + \bar \partial \bar c$ vanishes.

The second term of eq. \eqref{eq:hamiltonian-midpoint} is trivially a total $s$ derivative. The last term of eq. \eqref{eq:hamiltonian-midpoint} is also a total $s$ derivative. We can write it as
\begin{align}
    \int_0^1 ds \tr_{i p_0\sigma} \bigg( \frac{\partial \Psi(s)}{\partial s} \Psi^2 (s) \bigg) = & ~ \int_0^1 ds \frac{1}{3} \frac{\partial}{\partial s} \tr_{i p_0\sigma} ( \Psi^3(s) ) = \frac{1}{3} \tr_{i p_0\sigma} (\Psi^3).
    \label{eq:localized-interaction}
\end{align}
The first term of eq. \eqref{eq:hamiltonian-midpoint} can be split in two and reorganized as a total $s$ derivative and a term with $i p_0 Q \sigma$ insertion,
\begin{align}
    \tr_{Q\sigma} \bigg( \frac{\partial \Psi(s)}{\partial s} i p_0 \Psi(s) \bigg) = & ~ \frac{1}{2} \tr_{Q\sigma} \bigg( i p_0 \bigg( \frac{\partial \Psi(s)}{\partial s} \Psi(s) \bigg) - \bigg( i p_0 \frac{\partial \Psi(s)}{\partial s} \bigg) \Psi(s) \bigg) \nn
    & ~ + \frac{1}{2} \tr_{Q\sigma} \bigg( \frac{\partial \Psi(s)}{\partial s} i p_0 \Psi(s) \bigg) \nn
    = & ~ - \frac{1}{2} \tr_{i p_0Q\sigma} \bigg( \frac{\partial \Psi(s)}{\partial s} \Psi(s) \bigg) + \frac{1}{2} \frac{\partial}{\partial s} \tr_{Q\sigma} ( \Psi(s) i p_0 \Psi(s) ).
    \label{eq:trace-identity-p0}
\end{align}
In the first equality we split the term in two and use the Leibniz rule for $i p_0$. In the second equality we deform the contour of the total $i p_0$ derivative and make it act on the sigmoid, and recognize the remaining terms as a total derivative in $s$, after using cyclicity.

The third term in eq. \eqref{eq:hamiltonian-midpoint} is treated similarly: we split it in two, use Leibniz rule for $Q$ and gather total $Q$ and $s$ derivatives,
\begin{align}
    \tr_{i p_0\sigma} \bigg( \frac{\partial \Psi(s)}{\partial s} Q \Psi(s) \bigg) = & ~ \frac{1}{2} \tr_{i Q p_0\sigma} \bigg( \frac{\partial \Psi(s)}{\partial s} \Psi(s) \bigg) + \frac{1}{2} \frac{\partial}{\partial s} \tr_{i p_0\sigma} ( \Psi(s) Q \Psi(s) ).
    \label{eq:trace-identity-Q}
\end{align}
The difference in sign with respect to eq. \eqref{eq:trace-identity-p0} comes from the difference of Grassmann parity of $i p_0$ and $Q$. Summing eqs. \eqref{eq:trace-identity-p0} and \eqref{eq:trace-identity-Q}, the terms with the $i p_0 Q \sigma$ midpoint insertion cancel (because $[ip_0,Q]=0$ in the bulk).  Integrating over $s$ and summing all contributions, we find the Hamiltonian
\begin{align}
    H = & ~ \frac{1}{2} \tr_{Q\sigma} ( \Psi i p_0 \Psi ) + \frac{1}{2\pi i} \tr_{Q\sigma} ( \Psi ( c\partial X^0 - \bar c \bar \partial X^0 ) ) + \frac{1}{2} \tr_{i p_0\sigma} ( \Psi Q \Psi ) + \frac{1}{3} \tr_{i p_0\sigma} (\Psi^3).
\end{align}
At this point we can see that for a static solution, with $i p_0 \Psi = 0$, around a time-independent BCFT, when $c\partial X^0 - \bar c \bar \partial X^0 = 0$, the Hamiltonian evaluates to minus the action with the midpoint insertion of $i p_0\sigma$, which localizes the integral to a Cauchy slice. Hence the Hamiltonian of static solutions around static backgrounds equals minus the on-shell Lagrangian. Now we can use the equations of motion and combine the last two terms,
\begin{align}
    H = & ~ \frac{1}{2} \tr_{Q\sigma} (\Psi i p_0 \Psi) + \frac{1}{2\pi i} \tr_{Q\sigma} ( \Psi (c\partial X^0 - \bar c \bar \partial X^0) ) + \frac{1}{6} \tr_{i p_0 \sigma} (\Psi Q \Psi).
    \label{eq:hamiltonian-local-form}
\end{align}
We will evaluate this expression on the Erler-Maccaferri solution in the next subsection.

\subsection{Energy of the Erler-Maccaferri solution}
\label{sec:hamiltonian-derivation}

The expression of the Hamiltonian \eqref{eq:hamiltonian-local-form} can be further simplified when evaluated on the Erler-Maccaferri solution \eqref{eq:em-solution}. We start by splitting eq. \eqref{eq:hamiltonian-local-form} into three terms
\begin{align}
    H = & ~ H_1 + H_2 + H_3,
\end{align}
with
\begin{align}
    H_1 = & ~ \frac{1}{6} \text{Tr}_{i p_0 \sigma} ( \Psi Q \Psi ), \tab
    H_2 =  \frac{1}{2} \text{Tr}_{Q \sigma} ( \Psi i p_0 \Psi ), \tab
    H_3 =  \frac{1}{2\pi i} \text{Tr}_{Q\sigma} ( \Psi (c\partial X^0 - \bar c \bar \partial X^0) ).
    \label{eq:h1-h2}
\end{align}
Substituting eq. \eqref{eq:em-solution} in $H_1$, we find
\begin{align}
    H_1 = & ~ \frac{1}{6} \text{Tr}_{i p_0 \sigma} \Big( ( \tv - \Sigma \tv \bar \Sigma ) Q ( \tv - \Sigma \tv \bar \Sigma ) \Big) \nn
    = & ~ \frac{1}{6} \text{Tr}_{i p_0 \sigma} \Big( \tv Q \tv - \Sigma \tv \bar \Sigma Q \tv - \tv Q ( \Sigma \tv \bar \Sigma ) + \Sigma \tv \bar \Sigma Q ( \Sigma \tv \bar \Sigma ) \Big).
\end{align}
It is straightforward to check that
\begin{align}
    Q ( \Sigma \tv \bar \Sigma ) = & ~ - \tv \Sigma \tv \bar \Sigma + \Sigma \tv^2 \bar \Sigma - \Sigma \tv \bar \Sigma \tv.
\end{align}
Using this identity and also $Q\tv = -\tv^2$ we get
\begin{align}
    H_1 = & ~ \frac{1}{6} \text{Tr}_{i p_0\sigma} \Big( - \tv^3 + \Sigma \tv \bar \Sigma \tv^2 + \tv^2 \Sigma \tv \bar \Sigma - \tv \Sigma \tv^2 \bar \Sigma + \tv \Sigma \tv \bar \Sigma \tv \nn
    & ~ \hspace{2cm} - \Sigma \tv \bar \Sigma \tv \Sigma \tv \bar \Sigma + \Sigma \tv^3 \bar \Sigma - \Sigma \tv^2 \bar \Sigma \tv \Big).
\end{align}
Many of these terms can be combined using cyclicity. The trace is understood to be a correlator of the reference BCFT, so we can freely move $\tv$ from one side to the other, taking the proper signs into account. However, if we move $\Sigma$ or $\bar \Sigma$, the string field under the trace becomes an element of the Hilbert space of the target BCFT. To make the distinction whether the trace should be evaluated as a correlator on the target BCFT rather than on the reference one, we put a star superscript on the trace,
\begin{align}
    \text{Tr} ( \Sigma \cdots \bar \Sigma ) = & ~ \text{Tr}^* ( \cdots).
\end{align}
Using this notation, we can write $H_1$ as
\begin{align}
    H_1 = & ~ - \text{Tr}_{i p_0\sigma} \bigg( \frac{1}{6} \tv^3 - \frac{1}{2} \tv^2 \Sigma \tv \bar \Sigma \bigg) + \text{Tr}^*_{i p_0\sigma} \bigg( \frac{1}{6} \tv^3 - \frac{1}{2} \tv^2 \bar \Sigma \tv \Sigma \bigg).
\end{align}
We further note that the terms involving intertwining fields are $Q$-exact,
\begin{align}
    \text{Tr}_{i p_0\sigma} \Big( Q( \tv \Sigma \tv \bar \Sigma ) \Big) = & ~ - \tr^*_{i p_0\sigma} ( \tv^2 \bar \Sigma \tv \Sigma ) + \tr_{i p_0 \sigma} ( \tv^2 \Sigma \tv \bar \Sigma ).
\end{align}
The integration contour of $Q$ can be deformed to act on the midpoint, so we can write $H_1$ as
\begin{align}
    H_1 = & ~ - \frac{1}{6} \tr_{i p_0\sigma} (\tv^3) + \frac{1}{6} \tr^*_{i p_0\sigma} (\tv^3) - \frac{1}{2} \tr_{i Q p_0 \sigma} ( \tv \Sigma \tv \bar \Sigma ).
    \label{eq:h1}
\end{align}

Now we turn to another contribution of eq. \eqref{eq:h1-h2}, $H_2$. Substituting the Erler-Maccaferri solution, we find
\begin{align}
    H_2 = & ~ \frac{1}{2} \tr_{Q\sigma} ( \tv ip_0 \tv - \tv i p_0 (\Sigma \tv \bar \Sigma) - \Sigma \tv \bar \Sigma i p_0 \tv + \Sigma \tv \bar \Sigma ip_0 (\Sigma \tv \bar \Sigma) ).
\end{align}
The second and third terms can be reorganized using cyclicity and integrating $p_0$ by parts in one of the terms,
\begin{align}
    - \frac{1}{2} \tr_{Q\sigma} ( \tv i p_0 (\Sigma \tv \bar \Sigma) + \Sigma \tv \bar \Sigma i p_0 \tv ) = & ~ - \frac{1}{2} \tr_{Q\sigma} ( \tv i p_0 (\Sigma \tv \bar \Sigma) - (i p_0 \tv) \Sigma \tv \bar \Sigma ) \nn
    = & ~ \frac{1}{2} \tr_{ip_0 Q \sigma} ( \tv \Sigma \tv \bar \Sigma ) + \tr_{Q\sigma} ( (ip_0 \tv) \Sigma \tv \bar \Sigma ).
\end{align}
So we can write $H_2$ as
\begin{align}
    H_2 = & ~ \frac{1}{2} \tr_{Q\sigma} ( \tv ip_0 \tv + 2 (ip_0\tv) \Sigma \tv \bar \Sigma + \Sigma \tv \bar \Sigma ip_0 (\Sigma \tv \bar \Sigma) ) + \frac{1}{2} \tr_{ip_0Q\sigma} ( \tv \Sigma \tv \bar \Sigma ).
\end{align}
We now focus on the term with $ip_0 (\Sigma \tv \bar \Sigma)$. To find a simplification, we use the Leibniz rule for $ip_0$,
\begin{align}
   ip_0 ( \Sigma \tv \bar \Sigma) = & ~ (ip_0 \Sigma) \tv \bar \Sigma + \Sigma (ip_0 \tv) \bar \Sigma + \Sigma \tv (ip_0 \bar \Sigma).
\end{align}
Notice that $i p_0$ can act differently on different BCFTs. We can rearrange the terms using cyclicity, switching the trace to the target BCFT,
\begin{align}
    \frac{1}{2} \tr_{Q\sigma} \Big( \Sigma \tv \bar  \Sigma ip_0 ( \Sigma \tv \bar \Sigma) \Big) = & ~ \frac{1}{2} \tr^*_{Q\sigma} ( \tv ip_0 \tv ) - \frac{1}{2} \tr^*_{Q\sigma} ( \tv^2 \bar \Sigma (ip_0 \Sigma) ) \nn
    & ~ \hspace{4cm} + \frac{1}{2} \tr^*_{Q\sigma} ( \tv^2 (ip_0 \bar \Sigma) \Sigma ) .
\end{align}
The first term does not depend on the intertwining fields. The last two terms can be combined using $0 = ip_0 (\bar \Sigma \Sigma) = (ip_0 \bar \Sigma) \Sigma + \bar \Sigma (ip_0 \Sigma)$. To deal with them, we use the equation of motion of the tachyon vacuum in the form $Q_{\tv} \tv = \tv^2$ and BPZ conjugate $Q_{\tv}$,
\begin{align}
    \tr_{Q\sigma}^* ( \tv^2 \bar \Sigma ( ip_0 \Sigma ) ) = & ~ \tr_{Q\sigma}^* (Q_{\tv} \tv \bar \Sigma (ip_0 \Sigma) ) = \tr_{Q\sigma}^* \Big(\tv Q_{\tv} ( \bar \Sigma (ip_0 \Sigma) ) \Big)
\end{align}
Now we can use $Q_{\tv} \Sigma = Q_{\tv} \bar \Sigma = 0$, while keeping track of the commutator $[Q_{\tv},ip_0]$. We have
\begin{align}
    [Q_{\tv}, ip_0] \Sigma = & ~ [Q,ip_0] \Sigma + [[\tv,\cdot], ip_0] \Sigma \nn
    = & ~ [Q,ip_0] \Sigma + [\Sigma, ip_0 \tv].
\end{align}
So we find
\begin{align}
    \tr_{Q\sigma}^* ( \tv^2 \bar \Sigma (ip_0 \Sigma) ) = & ~ \tr_{Q\sigma}^* ( \tv \bar \Sigma [Q_{\tv}, ip_0] \Sigma ) \nn
    = & ~ \tr_{Q\sigma}^* ( \tv \bar \Sigma [Q,ip_0] \Sigma + \tv \bar \Sigma \Sigma ip_0 \tv - \tv \bar \Sigma ip_0\tv \Sigma ).
\end{align}
Evaluating $[Q,ip_0]$ at the endpoints of the contour, as in Section \ref{sec:Witten}, and using $\bar \Sigma \Sigma = 1$, we get
\begin{align}
    \tr_{Q\sigma}^* ( \tv^2 \bar \Sigma (ip_0 \Sigma) ) = & ~ \frac{1}{2\pi i} \tr_{Q\sigma}^* ( \tv ( c\partial X^0 - \bar c \bar \partial X^0 ) ) - \frac{1}{2\pi i} \tr_{Q\sigma}^* ( \tv \bar \Sigma ( c\partial X^0 - \bar c \bar \partial X^0 ) \Sigma ) \nn
    & ~ + \tr_{Q\sigma}^* ( \tv ip_0 \tv) + \tr_{Q\sigma} ( (ip_0\tv) \Sigma \tv \bar \Sigma ).
\end{align}
So $H_2$ becomes
\begin{align}
    H_2 = & ~ - \frac{1}{2} \tr_{Q\sigma}^* ( \tv ip_0 \tv) + \frac{1}{2} \tr_{Q\sigma} ( \tv ip_0 \tv) + \frac{1}{2} \tr_{ip_0Q\sigma} ( \tv \Sigma \tv \bar \Sigma ) \nn
    & ~ - \frac{1}{2\pi i} \tr_{Q\sigma}^* ( \tv ( c\partial X^0 - \bar c \bar \partial X^0 ) ) + \frac{1}{2\pi i} \tr_{Q\sigma}^* ( \tv \bar \Sigma ( c\partial X^0 - \bar c \bar \partial X^0 ) \Sigma ).
    \label{eq:h2}
\end{align}

And $H_3$ takes the form
\begin{align}
    H_3 = & ~ \frac{1}{2\pi i} \tr_{Q\sigma} \Big( ( \tv - \Sigma \tv \bar \Sigma ) ( c\partial X^0 - \bar c \bar \partial X^0 ) \Big) \nn
    = & ~ \frac{1}{2\pi i} \tr_{Q\sigma} \Big( \tv ( c\partial X^0 - \bar c \bar \partial X^0 ) \Big) - \frac{1}{2\pi i} \tr_{Q\sigma}^* \Big( \tv \bar \Sigma ( c\partial X^0 - \bar c \bar \partial X^0 ) \Sigma \Big),
    \label{eq:h3}
\end{align}
where in the second line we conjugate one of the terms to the target BCFT. Summing equations \eqref{eq:h1}, \eqref{eq:h2} and \eqref{eq:h3} we get the Hamiltonian
\begin{align}
    H = & ~ - \frac{1}{2} \tr_{Q\sigma}^* ( \tv ip_0 \tv ) - \frac{1}{2\pi i} \tr_{Q\sigma}^* ( \tv (c\partial X^0 - \bar c \bar \partial X^0) ) - \frac{1}{6} \tr^*_{ip_0\sigma} (\tv Q \tv) \nn
    & ~ + \frac{1}{2} \tr_{Q\sigma} ( \tv ip_0 \tv ) + \frac{1}{2\pi i} \tr_{Q\sigma} ( \tv (c\partial X^0 - \bar c \bar \partial X^0) ) + \frac{1}{6} \tr_{ip_0\sigma} (\tv Q \tv).
    \label{eq:hamiltonian-em}
\end{align}
The Hamiltonian has the expected form: it is the difference of correlators in the target and reference BCFTs.

Suppose we want to evaluate the energy of a solution around a time-independent BCFT. Then the first two terms of the second line in \eqref{eq:hamiltonian-em} vanish. For computational convenience, we chose to write $\tv$ with the $1/(1+K)$ on its right end, and now we can see why. In the second term of the first line in \eqref{eq:hamiltonian-em}, the flag and anti-flag state have cancelled, so the surface is a cylinder. This means that all the $c$ ghosts can be placed on the boundary. We can use the doubling trick and set $\bar c = c$. With the simple tachyon vacuum with $1/(1+K)$ on the right, cyclicity brings together the two fields $c$, and the term vanishes due to $c^2 = 0$. So we arrive at the expression
\begin{align}
    H = & ~ - \frac{1}{2} \tr_{Q\sigma}^* ( \tv ip_0 \tv ) - \frac{1}{6} \tr^*_{ip_0\sigma} (\tv Q \tv) + \frac{1}{6} \tr_{ip_0\sigma} (\tv Q \tv),
    \label{eq:hamiltonian-time-independent}
\end{align}
valid for the choice of writing the tachyon vacuum as $\tv = c (1+K) Bc/ (1+K)$. It holds for any Erler-Maccaferri solution (around a time-independent background) that does not deform the $X^1$ and ghost sectors. To evaluate it, one must know how $ip_0$ acts on tachyon vacuum solution $\tv$ of the target BCFT. The solution $\tv$ is expressed in terms of the string field $1/(1+K)$, which is in turn expressed as a Laplace transform of wedge states. So one must know how to act with $ip_0$ along a segment of the boundary of the target BCFT. We will see how this works in practice by computing the energy of the rolling tachyon solution in Section \ref{sec:rolling}.

\section{Rolling tachyon energy}
\label{sec:rolling}

We will now evaluate eq. \eqref{eq:hamiltonian-em} on the rolling tachyon solution generated by the marginal boundary deformation
\begin{align}
    \exp \bigg( \int dt \lambda \cosh X^0(t)\bigg),
    \label{eq:boundary-deformation}
\end{align}
around a static D$p$-brane configuration. The marginal operator $\cosh X^0 (t)$ has a singular OPE with itself, so eq. \eqref{eq:boundary-deformation} is understood to be under some normal ordering that avoids the potential singularities. The construction of the Erler-Maccaferri solution involves the flag states, but the Hamiltonian does not depend on these surface states explicitly, so we do not need to work out the geometry of these surface states.

\subsection{Rolling tachyon BCFT}

The information that we need in order to evaluate eq. \eqref{eq:hamiltonian-em} is how to compute correlation functions on the rolling tachyon BCFT, and how to act with $ip_0$ on a segment of the boundary. In \cite{Recknagel-Schomerus} the BCFT of a chiral marginal deformation is defined in terms of the associated boundary state.

Let the chiral marginal operator be a current $j^1$. Let $j^2$ and $j^3$ be other currents that form an $SU(2)$ algebra along with $j^1$, satisfying the OPE
\begin{align}
    j^a (z) j^b (w) \sim & ~ \frac{1}{2} \frac{\delta^{ab}}{(z-w)^2} + \frac{i\epsilon^{abc} j^c (w)}{z-w}.
    \label{eq:su2-algebra}
\end{align}
Along with the chiral currents $j^a, a=1,2,3$, we assume the existence of the anti-holomorphic counterparts $\bar j^a, a=1,2,3$. In the reference BCFT, the gluing conditions on the boundary identify each current with its corresponding anti-holomorphic one,
\begin{align}
    j^1 = \bar j^1, \tab
    j^2 = \bar j^2, \tab
    j^3 = \bar j^3.
\end{align}
On the BCFT deformed by the marginal operator $\lambda j^1$, these gluing conditions are rotated around the $a=1$ direction with angle $2\pi\lambda$, as
\begin{align}
    j^1 = & ~ \bar j^1, \nn
    j^2 = & ~ \bar j^2 \cos (2\pi\lambda) - \bar j^3 \sin(2\pi \lambda), \nn
    j^3 = & ~ \bar j^2 \sin(2\pi \lambda) + \bar j^3 \cos(2\pi \lambda).
    \label{eq:gluing-conditions}
\end{align}
The construction in \cite{Recknagel-Schomerus} requires that the BCFT has Euclidean signature in spacetime in order for the boundary state computation to be well-defined. However, we want to avoid considering a string field theory solution in Euclidean signature, in order to keep a clear interpretation of the time direction in the construction of the Hamiltonian. We will instead analytically continue the gluing conditions \eqref{eq:gluing-conditions} for a BCFT in Lorentzian spacetime signature.

We want to describe the rolling tachyon solution, whose corresponding chiral marginal deformation is $\cosh X^0$. We consider a set of chiral currents,
\begin{align}
    j^1 = \cosh X^0, \tab
    j^2 = - i \sinh X^0, \tab
    j^3 = \partial X^0.
\end{align}
The currents satisfy the same algebra as eq. \eqref{eq:su2-algebra}. Therefore the gluing conditions of the rolling tachyon BCFT take the same form as eq. \eqref{eq:gluing-conditions}. We also consider the anti-chiral currents with the same structure.

To evaluate eq. \eqref{eq:hamiltonian-time-independent} we must know how to act with $ip_0$ on the string fields. We can write it in terms of $j^3$,
\begin{align}
    ip_0 = & ~ \oint \frac{dz}{2\pi i} j^3 (z) - \oint \frac{d\bar z}{2\pi i} \bar j^3 (\bar z).
\end{align}
Its action on a boundary segment is given by an integral along the boundary,
\begin{align}
    ip_0 = & ~ \int \frac{dx}{2\pi i} ( j^3 - \bar j^3 ).
\end{align}
Using the gluing conditions, we reexpress both $j^3$ and $\bar j^3$, as
\begin{align}
    j^3 = & ~ \bar j^2 \sin (2\pi\lambda) + \bar j^3 \cos(2\pi\lambda), \tab
    \bar j^3 = - j^2 \sin (2\pi\lambda) + j^3 \cos(2\pi\lambda).
\end{align}
So $ip_0$ takes the form
\begin{align}
    ip_0 = & ~ \int \frac{dx}{2\pi i} \Big( (j^2+\bar j^2) \sin(2\pi\lambda) - (j^3-\bar j^3) \cos(2\pi\lambda) \Big) \nn
    = & ~ \int \frac{dx}{2\pi i} \mathbb{J}^2 \sin(2\pi\lambda) - ip_0 \cos(2\pi\lambda),
\end{align}
where we define
\begin{align}
    \mathbb{J}^2 = j^2 + \bar j^2.
\end{align}
The action of $ip_0$ is the insertion of $\mathbb{J}^2$ with a normalization,
\begin{align}
    ip_0 = & ~ \frac{\sin(2\pi\lambda)}{1+\cos(2\pi\lambda)} \int \frac{dx}{2\pi i} \mathbb{J}^2 \nn
    = & ~ \tan(\pi\lambda) \int \frac{dx}{2\pi i} \mathbb{J}^2.
    \label{eq:p0-action}
\end{align}
Notice that the action of $ip_0$ is singular for $\lambda=1/2$, which is known to be a special point in the space of solutions \cite{GIR}. If $\Omega^\alpha$ is a wedge state on the rolling tachyon BCFT, then the action of $ip_0$ is given by
\begin{align}
    ip_0 \Omega^\alpha = & ~ - \frac{\tan(\pi\lambda)}{2\pi i} \int_0^\alpha d\beta \Omega^\beta \mathbb{J}^2 \Omega^{\alpha-\beta}.
    \label{eq:p0-string-field}
\end{align}
We use the left handed convention \cite{Erler:AnalyticLectures} when writing correlators, so if the integral of eq. \eqref{eq:p0-action} runs from left to right in the UHP, the integral in eq. \eqref{eq:p0-string-field} would run from $\alpha$ to $0$, which explains the sign flip. From this we find the action of $ip_0$ on $K$,
\begin{align}
    ip_0 K = & ~ - ip_0 \frac{\partial}{\partial \alpha} \Omega^\alpha \Big|_{\alpha=0} = \frac{\tan(\pi\lambda)}{2\pi i} \mathbb{J}^2.
\end{align}
The derivative $\partial/\partial \alpha$ acts on both the upper limit of the integral in \eqref{eq:p0-string-field} and on the integrand. The second contribution reduces to an integral over an interval of vanishing length when we set $\alpha=0$. We also act on $1/(1+K)$,
\begin{align}
    ip_0 \frac{1}{1+K} = & ~ ip_0 \int_0^\infty dt e^{-t} \Omega^t \nn
    = & ~ - \frac{\tan(\pi\lambda)}{2\pi i} \int_0^\infty dt \int_0^t ds e^{-t} \Omega^s \mathbb{J}^2 \Omega^{t-s} \nn
    = & ~ - \frac{\tan(\pi\lambda)}{2\pi i} \int_0^\infty dt_1 \int_0^\infty dt_2 e^{-t_1-t_2} \Omega^{t_1} \mathbb{J}^2 \Omega^{t_2} \nn
    = & ~ - \frac{\tan(\pi\lambda)}{2\pi i} \frac{1}{1+K} \mathbb{J}^2 \frac{1}{1+K}.
\end{align}
Consequently, we know how to act on the tachyon vacuum,
\begin{align}
    ip_0 \tv = & ~ ip_0 \bigg( c(1+K)Bc \frac{1}{1+K} \bigg) = c (ip_0K) Bc \frac{1}{1+K} + c(1+K) Bc ip_0 \frac{1}{1+K} \nn
    = & ~ \frac{\tan(\pi\lambda)}{2\pi i} \bigg( c \mathbb{J}^2 Bc \frac{1}{1+K} - c(1+K)Bc \frac{1}{1+K} \mathbb{J}^2 \frac{1}{1+K} \bigg).
    \label{eq:p0-tv}
\end{align}
We can simplify the first and second terms by anticommuting $B$ and $c$,
\begin{align}
    c \mathbb{J}^2 Bc \frac{1}{1+K} = & ~ c \mathbb{J}^2 \frac{1}{1+K}, \nn
    c(1+K) Bc \frac{1}{1+K} \mathbb{J}^2 \frac{1}{1+K} = & ~ \bigg( c - cKc \frac{B}{1+K} \bigg) \mathbb{J}^2 \frac{1}{1+K}.
\end{align}
We used the property $c^2=0$ to make some terms vanish. Then eq. \eqref{eq:p0-tv} reduces to
\begin{align}
    ip_0 \tv = & ~ \frac{\tan(\pi\lambda)}{2\pi i} c K c \frac{B}{1+K} \mathbb{J}^2 \frac{1}{1+K}.
    \label{eq:p0-action-2}
\end{align}
With this identity at hand, we proceed to evaluate the Hamiltonian.

\subsection{Evaluation of the energy}

To compute the energy of the rolling tachyon solution, we take the Hamiltonian of eq. \eqref{eq:hamiltonian-time-independent}, valid for solutions around a static background, and evaluate it on the rolling tachyon BCFT.

In the $X^0$ sector, we must specify how correlators of the Lorentzian rolling tachyon BCFT are evaluated. We follow the prescription of \cite{Bernardes:2026RollingTachyons}: each correlator is expanded in plane waves $e^{\pm n X^0}$. For the terms with zero total energy, the zero-mode integral produces the energy-conserving delta function, $\langle e^{iEX^0}\rangle = 2\pi \delta(E)$, normalized by the time volume as in the reference BCFT. The terms with nonzero imaginary total energy are discarded.

We first show that the terms with the $ip_0\sigma$ midpoint insertion cancel between the reference and the target BCFT. The simple tachyon vacuum can be written in the form
\begin{align}
    \tv = c \frac{1}{1+K} - Q \bigg( c \frac{B}{1+K} \bigg).
\end{align}
The contribution from the reference BCFT is
\begin{align}
    \frac{1}{6} \tr_{ip_0\sigma} (\tv Q \tv)
    = & ~ \frac{1}{6} \tr_{ip_0\sigma} \bigg( \bigg( c \frac{1}{1+K} - Q \bigg( c\frac{B}{1+K} \bigg) \bigg) Q \bigg( c \frac{1}{1+K} \bigg) \bigg) \nn
    = & ~ \frac{1}{6} \tr_{ip_0\sigma} \bigg( c \frac{1}{1+K} c \partial c \frac{1}{1+K} \bigg) - \frac{1}{6} \tr_{ip_0\sigma} \bigg( Q \bigg( c \frac{B}{1+K} \bigg) Q \bigg( c \frac{1}{1+K} \bigg) \bigg).
    \label{eq:brane-tension-setup}
\end{align}
In the second term, we deform the contour of the first $Q$ to BPZ conjugate it. It acts on the other $Q$, giving $Q^2=0$, so that term vanishes. And it also acts on the midpoint insertion, changing it to
\begin{align}
    iQp_0 \sigma = & ~ \int_{-\infty}^\infty \frac{dE}{2\pi} \dot \sigma(E) i E \no{ ( c(i\infty) \partial X^+ (i\infty) + \bar c (\bar i \infty) \bar \partial X^+ (\bar i \infty ) ) e^{iEX^+(i\infty,\bar i\infty)} }.
\end{align}
This term becomes a correlator with a single $\partial X^+$ or $\bar \partial X^+$ insertion, which vanishes. We turn to the other term, where we can express the $1/(1+K)$ as a Laplace transform of wedge states
\begin{align}
    \frac{1}{1+K} = & ~ \int_0^\infty dt e^{-t} \Omega^t.
\end{align}
The expression becomes
\begin{align}
    \frac{1}{6} \tr_{ip_0\sigma} (\tv Q \tv) = & ~ \frac{1}{6} \int_0^\infty dt_1 \int_0^\infty dt_2 e^{-t_1-t_2} \tr_{ip_0\sigma} ( c\Omega^{t_1} c \partial c \Omega^{t_2} ) \nn
    = & ~ \frac{1}{6} \int_{-\infty}^\infty \frac{dE}{2\pi} \dot \sigma(E) \int_0^\infty dt_1 \int_0^\infty dt_2 e^{-t_1-t_2} \nn
    & ~ \hspace{2cm} \langle e^{iEX^+(i\infty,\bar i\infty)} c(t_1+t_2) c(t_2) \partial c (t_2) \rangle_{C_{t_1+t_2}},
\end{align}
where we expressed the trace as a correlator on the cylinder frame $C_{t_1+t_2}$ of circumference $t_1+t_2$. We go to the UHP with the map
\begin{align}
    C_L^{-1} (z) = & ~ \tan \bigg( \frac{\pi z}{L} \bigg),
\end{align}
for $L=t_1+t_2$. The plane wave has conformal weight zero, so only the ghost insertions carry conformal factors in the transformation. The correlator in the Neumann sector generates a delta function $2\pi \delta(E)$, which sets $E=0$ after performing the integral over $E$. The remaining matter correlators in the Neumann sector supply the volume factor $V_p = (2\pi)^{p} \delta^{p} (0)$. Using $\dot \sigma(E=0)=1$, the correlator evaluates to
\begin{align}
    \frac{1}{6} \tr_{ip_0\sigma} (\tv Q \tv) = & ~ - \frac{V_p}{6} \int_0^\infty dt_1 \int_0^\infty dt_2 e^{-t_1-t_2} \frac{( C_{t_1+t_2}^{-1} (t_1+t_2) - C_{t_1+t_2}^{-1} (t_2) )^2}{(C_{t_1+t_2}^{-1})' (t_1+t_2) (C_{t_1+t_2}^{-1})'(t_2)} \nn
    = & ~ - \frac{V_p}{6}  \int_0^\infty dt_1 \int_0^\infty dt_2 e^{-t_1-t_2} \frac{(t_1+t_2)^2}{\pi^2} \sin^2 \bigg( \frac{\pi t_1}{t_1+t_2} \bigg) \nn
    = & ~ - \frac{V_p}{2\pi^2} = - V_p T_p.
\end{align}
In the target BCFT the structure is the same: the correlator in the $X^1$ sector generates the delta function that fixes $E=0$, then the correlator in the $X^0$ sector generates the same volume of time as in the reference BCFT, following our prescription. We thus have the cancellation
\begin{align}
    \frac{1}{6} \tr_{ip_0\sigma} ( \tv Q \tv ) - \frac{1}{6} \tr^*_{ip_0\sigma} (\tv Q \tv) = 0,
\end{align}
leaving
\begin{align}
    H = & ~ - \frac{1}{2} \tr^*_{Q\sigma} ( \tv ip_0 \tv )
\end{align}
as the only contribution to the Hamiltonian.

Now we use the expression \eqref{eq:p0-action-2} to put the Hamiltonian in the form
\begin{align}
    H = & ~ - \frac{\tan(\pi\lambda)}{4\pi i} \tr_{Q\sigma}^* \bigg( c(1+K) Bc \frac{1}{1+K} ~ cKc \frac{B}{1+K} \mathbb{J}^2 \frac{1}{1+K} \bigg).
\end{align}
Again we use $[B,c]=1$ to shift the position of the first $B$ and second $c$ inside the trace, so the expression becomes
\begin{align}
    H = & ~ \frac{\tan(\pi\lambda)}{4\pi i} \tr_{Q\sigma}^* \bigg( cKc \frac{B}{1+K} cKc \frac{B}{1+K} \mathbb{J}^2 \frac{1}{1+K} \bigg).
\end{align}
Then we bring the second $B$ to the left, anticommuting it with the $c$ fields, until it hits the other $B$, so we can use $B^2=0$. We get
\begin{align}
    H = & ~ - \frac{\tan(\pi\lambda)}{4\pi i} \tr_{Q\sigma}^* \bigg( - c K c \frac{B}{1+K} c \frac{K}{1+K} \mathbb{J}^2 \frac{1}{1+K} + cKc \frac{BK}{1+K} c \frac{1}{1+K} \mathbb{J}^2 \frac{1}{1+K} \bigg).
\end{align}
We can use the property
\begin{align}
    \frac{K}{1+K} = 1 - \frac{1}{1+K}
    \label{eq:K-identity}
\end{align}
in order to get
\begin{align}
    H = & ~ \frac{\tan(\pi\lambda)}{4\pi i} \tr_{Q\sigma}^* \bigg( cKc \frac{B}{1+K} c \mathbb{J}^2 \frac{1}{1+K} - cKc \frac{B}{1+K} c \frac{1}{1+K} \mathbb{J}^2 \frac{1}{1+K} \nn
    & ~ \hspace{3.5cm} - cKcBc \frac{1}{1+K} \mathbb{J}^2 \frac{1}{1+K} + cKc \frac{B}{1+K} c \frac{1}{1+K} \mathbb{J}^2 \frac{1}{1+K} \bigg).
\end{align}
The second and fourth terms cancel. In the third term we can use $cBc = c(1-cB) = c$. We find
\begin{align}
    H = & ~ \frac{\tan(\pi\lambda)}{4\pi i} \tr_{Q\sigma}^* \bigg( c K c \frac{1}{1+K} (Bc - 1) \mathbb{J}^2 \frac{1}{1+K} \bigg) \nn
    = & ~ -\frac{\tan(\pi\lambda)}{4\pi i} \tr_{Q\sigma}^* \bigg( cKc \frac{1}{1+K} c \mathbb{J}^2 \frac{B}{1+K} \bigg).
\end{align}
As a last step to simplify the expression, we recognize the $Q$-exact term  $cKc = Qc$, and deform the BRST contour to make it act on the remaining fields. The midpoint insertion is BRST closed, as well as $K$ and $c\mathbb{J}^2$. The only contribution comes from $QB = K$,
\begin{align}
    H = & ~ \frac{\tan(\pi\lambda)}{4\pi i} \tr_{Q\sigma}^* \bigg( c \frac{1}{1+K} c \mathbb{J}^2 \frac{K}{1+K} \bigg).
\end{align}
Once again using eq. \eqref{eq:K-identity}, as well as cyclicity, we get
\begin{align}
    H = & ~ \frac{\tan(\pi\lambda)}{4\pi i} \tr_{Q\sigma}^* \bigg( \frac{1}{1+K} c \mathbb{J}^2 \frac{1}{1+K} c \bigg).
    \label{eq:hamiltonian-kbc}
\end{align}
Now we write the expression as a correlator. We express the functions of $K$ as integral transforms of wedge states,
\begin{align}
    H = & ~ \frac{\tan(\pi\lambda)}{4\pi i} \int_0^\infty dt_1 \int_0^\infty dt_2 e^{-t_1-t_2} \tr_{Q\sigma}^* ( \Omega^{t_1} c \mathbb{J}^2 \Omega^{t_2} c ).
\end{align}
The trace can be directly expressed as a correlation function on a cylinder $C_{t_1+t_2}$ of circumference $t_1+t_2$,
\begin{align}
    H = & ~ \frac{\tan(\pi\lambda)}{4\pi i} \int_0^\infty dt_1 \int_0^\infty dt_2 e^{-t_1-t_2} \int_{-\infty}^\infty \frac{dE}{2\pi} \dot \sigma(E) \nn
    & ~ \hspace{4cm} \Big( \langle c(i\infty) \no{ \partial X^+ (i\infty) e^{iEX^+(i\infty,\bar i \infty)} } c(-t_2) \mathbb{J}^2 (-t_2) c(0) \rangle_{C_{t_1+t_2}}^* \nn
    & ~ \hspace{4.5cm} + \langle \bar c(\bar i\infty) \no{ \bar \partial X^+ (\bar i\infty) e^{iEX^+(i\infty,\bar i\infty)} } c(-t_2) \mathbb{J}^2 (-t_2) c(0) \rangle_{C_{t_1+t_2}}^* \Big).
    \label{eq:correlator-hamiltonian}
\end{align}
It is useful to change the integration variables to
\begin{align}
    t = t_1 + t_2, \tab \varphi = \frac{\pi t_2}{t_1+t_2}.
\end{align}
The integration measure becomes $dt_1 dt_2 = (t/\pi) dt d\varphi$ and the integration limits are $0\leq t \leq \infty$, $0 \leq \varphi \leq \pi$.  The midpoint insertions have conformal weight zero, as well as the $c\mathbb{J}^2$ insertion on the boundary. The conformal map satisfies $C^{-1}_t (0) = 0$ and $C^{-1}_t (-t_2) = -\tan \varphi$. The only conformal weight factor comes from $c(0)$, and it evaluates to $1/(C^{-1}_t)'(0)=t/\pi$. The Hamiltonian is expressed as
\begin{align}
    H = & ~ \frac{\tan(\pi\lambda)}{4\pi i} \int_0^\infty dt t^2 e^{-t} \int_0^\pi \frac{d\varphi}{\pi^2} \int_{-\infty}^\infty \frac{dE}{2\pi} \dot \sigma(E) \nn
    & ~ \hspace{4cm} \Big( \langle c(i) \no{ \partial X^+ (i) e^{iEX^+ (i,\bar i) } } c( - \tan \varphi ) \mathbb{J}^2 (- \tan \varphi) c(0) \rangle_\text{UHP}^* \nn
    & ~ \hspace{4.5cm} + \langle \bar c(\bar i) \no{ \bar \partial X^+ (\bar i) e^{iEX^+(i,\bar i)} } c( - \tan \varphi) \mathbb{J}^2 ( - \tan \varphi ) c(0) \rangle_\text{UHP}^* \Big).
\end{align}
Notice that the correlator does not depend on $t$, so we can perform the integral,
\begin{align}
    \int_0^\infty dt t^2 e^{-t} = \Gamma(3) = 2.
\end{align}
The correlators in the $X^1$ sector evaluate to
\begin{align}
    \langle e^{iEX^1(i,\bar i)} \rangle_\text{UHP} = & ~ 2\pi \delta(E), \nn
    \langle \no{ \partial X^1 (i) e^{iEX^1 (i,\bar i)} } \rangle_\text{UHP} = & ~ \langle \no{ \bar \partial X^1 (\bar i) e^{iEX^1 (i,\bar i)} } \rangle_\text{UHP} = 0.
\end{align}
We identify the currents $j^3 = \partial X^0$ and $\bar j^3 = \bar \partial X^0$, so $H$ is expressed as
\begin{align}
    H = & ~ \frac{\tan(\pi\lambda)}{2\pi i} \int_0^\pi \frac{d \varphi}{\pi^2} \bigg( \langle c(i) c ( \tan \varphi) c(0) \rangle_\text{UHP} \Big( \langle j^3(i) j^2 ( \tan \varphi) \rangle^{'*}_\text{UHP} + \langle j^3(i) \bar j^2 ( \tan \varphi) \rangle^{'*}_\text{UHP} \Big) \nn
    & ~ \hspace{3cm} + \langle \bar c(\bar i) c ( \tan \varphi) c(0) \rangle_\text{UHP} \Big( \langle \bar j^3 (\bar i) j^2 ( \tan \varphi ) \rangle^{'*}_\text{UHP} + \langle \bar j^3 (\bar i) \bar j^2 ( \tan \varphi ) \rangle^{'*}_\text{UHP} \Big) \bigg),
\end{align}
where $\langle \cdots \rangle^{'*}_\text{UHP}$ denotes the correlator with the $X^1$ sector removed. The ghost sector gives
\begin{align}
    \langle c(i) c(\tan \varphi) c(0) \rangle_\text{UHP} = & ~ i \tan \varphi (i-\tan \varphi), \tab
    \langle \bar c (\bar i) c (\tan \varphi) c(0) \rangle_\text{UHP} = i \tan \varphi (i + \tan \varphi),
\end{align}
and the correlators on the remaining spatial Neumann directions generate the volume factor $V_{p-1} = (2\pi)^{p-1} \delta^{p-1} (0)$. In the $X^0$ sector, the two point function of currents is fixed by conformal invariance,
\begin{align}
    \langle j^a (z) j^b (w) \rangle_\text{UHP}^* = \frac{V}{2} \frac{\delta^{ab}}{(z-w)^2},
\end{align}
where $V=2\pi \delta(0)$. In particular $\langle j^3 j^2\rangle^*$ vanishes, and the prescription enters only through the normalization $V$. Using the doubling trick with the gluing conditions \eqref{eq:gluing-conditions}, we have
\begin{align}
    \langle j^3 (i) j^2 (\tan \varphi) \rangle^*_\text{UHP} + \langle j^3 (i) \bar j^2 (\tan \varphi) \rangle^*_\text{UHP} = & ~ \frac{V}{2} \frac{1}{(i-\tan \varphi)^2} 
    \sin(2\pi\lambda), \nn
    \langle \bar j^3 (\bar i) j^2 (\tan \varphi) \rangle^*_\text{UHP} + \langle \bar j^3 (\bar i) \bar j^2 (\tan \varphi) \rangle^*_\text{UHP} = & ~ -\frac{V}{2} \frac{1}{(-i-\tan \varphi)^2} \sin(2\pi\lambda).
\end{align}
The Hamiltonian can thus be expressed as
\begin{align}
    H = & ~ V_p \frac{\tan(\pi\lambda)}{2\pi i} \int_0^\pi \frac{d \varphi}{\pi^2} \bigg( \frac{i}{2} \frac{\tan\varphi}{i-\tan\varphi} - \frac{i}{2} \frac{\tan\varphi}{i+\tan\varphi}  \bigg) \sin(2\pi\lambda)
    \nn
    = & ~ - \frac{V_p}{2\pi} \tan(\pi\lambda) \sin(2\pi\lambda) \int_0^\pi \frac{d\varphi}{\pi^2} \frac{\tan^2\varphi}{1+\tan^2\varphi},
\end{align}
with $V_p = (2\pi)^p \delta^p (0)$. Using $\tan(\pi\lambda) \sin(2\pi\lambda) = 2 \sin^2 (\pi\lambda)$, we write
\begin{align}
    H = & ~ - \frac{V_p }{\pi} \sin^2 (\pi\lambda) \underbrace{\int_0^\pi \frac{d\varphi}{\pi^2} \frac{\tan^2 \varphi}{1+\tan^2\varphi}}_{=1/(2\pi)} = - \frac{V_p}{2\pi^2} \sin^2 (\pi\lambda).
\end{align}
Identifying the D$p$-brane tension $T_p = 1/(2\pi^2)$, we find
\begin{align}
    H = & ~ - V_p T_p \sin^2 (\pi \lambda).
\end{align}
Adding the energy $V_p T_p$ of the reference D$p$-brane, we find the energy measured relative to the tachyon vacuum,
\begin{align}
    E = & ~ V_p T_p + H = V_p T_p \cos^2 (\pi \lambda),
\end{align}
which agrees with Sen's result \cite{Sen:2002RollingTachyon}.

\section{Discussion}

We have evaluated the energy as a conserved charge associated to time translation symmetry. This quantity expresses the notion of \emph{inertial energy}. It has been known how to express the energy of static solutions in the Witten OSFT in terms of the Ellwood invariant \cite{Ellwood,KSM,Baba-Ishibashi}, which defines the coupling of Witten OSFT to on-shell closed string backgrounds \cite{KOZ,MV}. This gives a notion of coupling open string field theory to gravity, so the \emph{gravitational energy} is expected to be expressed as an Ellwood invariant. It should be possible to express the Hamiltonian of the Witten OSFT in the form of an Ellwood invariant and show that the inertial and gravitational energies agree. We leave this question for future work.

\subsection*{Acknowledgments}

We thank Ted Erler, Atakan Hilmi Fırat, Martin Schnabl and Jakub Vošmera for discussions and comments on the draft. The research of VB  was co-funded by the European Union and supported by the Czech Ministry of Education, Youth and Sports (Project FORTE CZ.02.01.01/00/22 008/0004632). The research of RP was supported by the Grant Agency of the Czech Republic under the grant project 26-23188S.

\bibliography{refs.bib}

@article{Witten:OSFT,
  author       = {Witten, Edward},
  title        = {Noncommutative geometry and string field theory},
  journal      = {Nucl. Phys. B},
  year         = {1986},
  volume       = {268},
  pages        = {253--294},
  doi          = {10.1016/0550-3213(86)90155-0}
}

@article{Witten:SymplecticForm,
  author       = {Witten, Edward},
  title        = {Interacting field theory of open superstrings},
  journal      = {Nucl. Phys. B},
  year         = {1986},
  volume       = {276},
  pages        = {291--324},
  doi          = {10.1016/0550-3213(86)90298-1}
}

@article{Ellwood,
  author       = {Ellwood, Ian},
  title        = {The closed string tadpole in open string field theory},
  journal      = {JHEP},
  year         = {2008},
  volume       = {08},
  pages        = {063},
  doi          = {10.1088/1126-6708/2008/08/063},
  eprint       = {0804.1131},
  archivePrefix= {arXiv},
  primaryClass = {hep-th}
}

@article{MV,
  author       = {Maccaferri, Carlo and Vo{\v{s}}mera, Jakub},
  title        = {Closed string deformations in open string field theory. {P}art {I}. {B}osonic string},
  journal      = {JHEP},
  year         = {2021},
  volume       = {09},
  pages        = {047},
  doi          = {10.1007/JHEP09(2021)047},
  eprint       = {2103.04919},
  archivePrefix= {arXiv},
  primaryClass = {hep-th}
}

@article{Zwiebach:1992Interpolating,
  author       = {Zwiebach, Barton},
  title        = {Interpolating string field theories},
  journal      = {Mod. Phys. Lett. A},
  year         = {1992},
  volume       = {7},
  pages        = {1079--1090},
  doi          = {10.1142/S0217732392000951},
  eprint       = {hep-th/9202015},
  archivePrefix= {arXiv},
  primaryClass = {hep-th}
}

@article{ElectricField,
  author       = {Bernardes, Vin{\'i}cius and Erler, Theodore and F{\i}rat, Atakan Hilmi},
  title        = {Symplectic structure in open string field theory. {P}art {III}. {E}lectric field},
  journal      = {JHEP},
  year         = {2026},
  volume       = {06},
  pages        = {052},
  doi          = {10.1007/JHEP06(2026)052},
  eprint       = {2604.01273},
  archivePrefix= {arXiv},
  primaryClass = {hep-th}
}

@article{Bernardes:2026RollingTachyons,
  author       = {Bernardes, Vin{\'i}cius and Erler, Theodore and F{\i}rat, Atakan Hilmi},
  title        = {Symplectic structure in open string field theory. {P}art {I}. {R}olling tachyons},
  journal      = {JHEP},
  year         = {2026},
  volume       = {02},
  pages        = {063},
  doi          = {10.1007/JHEP02(2026)063},
  eprint       = {2511.03777},
  archivePrefix= {arXiv},
  primaryClass = {hep-th}
}

@article{Bernardes:Lump,
  author       = {Bernardes, Vin{\'i}cius and Erler, Theodore and F{\i}rat, Atakan Hilmi},
  title        = {Symplectic structure in open string field theory. {P}art {II}. {S}liding lump},
  journal      = {JHEP},
  year         = {2026},
  volume       = {02},
  pages        = {064},
  doi          = {10.1007/JHEP02(2026)064},
  eprint       = {2511.15781},
  archivePrefix= {arXiv},
  primaryClass = {hep-th}
}

@article{Bernardes:ConservedCharges,
  author       = {Bernardes, Vin{\'i}cius and Erler, Theodore and F{\i}rat, Atakan Hilmi},
  title        = {Conserved charges and {$L_\infty$} algebras},
  year         = {2026},
  eprint       = {2606.26224},
  archivePrefix= {arXiv},
  primaryClass = {hep-th}
}

@article{Pressure,
  author       = {Bernardes, Vin{\'i}cius and F{\i}rat, Atakan Hilmi},
  title        = {Pressure in {W}ilsonian effective field theory},
  year         = {2026},
  eprint       = {2610.30249},
  archivePrefix= {arXiv},
  primaryClass = {hep-th}
}

@article{CMY,
  author       = {Cho, Minjae and Mazel, Ben and Yin, Xi},
  title        = {Rolling tachyon and the phase space of open string field theory},
  journal      = {JHEP},
  year         = {2025},
  volume       = {04},
  pages        = {129},
  doi          = {10.1007/JHEP04(2025)129},
  eprint       = {2310.17895},
  archivePrefix= {arXiv},
  primaryClass = {hep-th}
}

@article{Bernardes:CovariantPhaseSpace,
  author       = {Bernardes, Vin{\'i}cius and Erler, Theodore and F{\i}rat, Atakan Hilmi},
  title        = {Covariant phase space and {$L_\infty$} algebras},
  journal      = {JHEP},
  year         = {2025},
  volume       = {09},
  pages        = {057},
  doi          = {10.1007/JHEP09(2025)057},
  eprint       = {2506.20706},
  archivePrefix= {arXiv},
  primaryClass = {hep-th}
}

@article{SenZwiebach:2024Review,
  author       = {Sen, Ashoke and Zwiebach, Barton},
  title        = {String Field Theory: A Review},
  year         = {2024},
  eprint       = {2405.19421},
  archivePrefix= {arXiv},
  primaryClass = {hep-th}
}

@article{Sen:2002RollingTachyon,
  author       = {Sen, Ashoke},
  title        = {Rolling tachyon},
  journal      = {JHEP},
  year         = {2002},
  volume       = {04},
  pages        = {048},
  doi          = {10.1088/1126-6708/2002/04/048},
  eprint       = {hep-th/0203211},
  archivePrefix= {arXiv},
  primaryClass = {hep-th}
}

@article{Moeller:2002vx,
  author       = {Moeller, Nicolas and Zwiebach, Barton},
  title        = {Dynamics with infinitely many time derivatives and rolling tachyons},
  journal      = {JHEP},
  year         = {2002},
  volume       = {10},
  pages        = {034},
  doi          = {10.1088/1126-6708/2002/10/034},
  eprint       = {hep-th/0207107},
  archivePrefix= {arXiv},
  primaryClass = {hep-th}
}

@article{Yang,
  author       = {Yang, Haitang},
  title        = {Stress tensors in $p$-adic string theory and truncated {OSFT}},
  journal      = {JHEP},
  year         = {2002},
  volume       = {11},
  pages        = {007},
  doi          = {10.1088/1126-6708/2002/11/007},
  eprint       = {hep-th/0209197},
  archivePrefix= {arXiv},
  primaryClass = {hep-th}
}

@article{StettingerAli,
  author       = {Ali, Mohd and Stettinger, Georg},
  title        = {The {BEF} symplectic form: A {L}agrangian perspective},
  year         = {2026},
  eprint       = {2604.07334},
  archivePrefix= {arXiv},
  primaryClass = {hep-th}
}

@article{Erler:AnalyticLectures,
    author        = "Erler, Theodore",
    title         = "{Four Lectures on Analytic Solutions in Open String Field Theory}",
    eprint        = "1912.00521",
    archivePrefix = "arXiv",
    primaryClass  = "hep-th",
    doi           = "10.1016/j.physrep.2022.06.004",
    journal       = "Phys. Rept.",
    volume        = "980",
    pages         = "1--95",
    year          = "2022"
}

@article{erler-schnabl-simple,
    author        = "Erler, Theodore and Schnabl, Martin",
    title         = "{A Simple Analytic Solution for Tachyon Condensation}",
    eprint        = "0906.0979",
    archivePrefix = "arXiv",
    primaryClass  = "hep-th",
    doi           = "10.1088/1126-6708/2009/10/066",
    journal       = "JHEP",
    volume        = "10",
    pages         = "066",
    year          = "2009"
}

@article{KOZ,
    author        = "Kiermaier, Michael and Okawa, Yuji and Zwiebach, Barton",
    title         = "{The boundary state from open string fields}",
    eprint        = "0810.1737",
    archivePrefix = "arXiv",
    primaryClass  = "hep-th",
    year          = "2008"
}

@article{erler-maccaferri,
    author        = "Erler, Theodore and Maccaferri, Carlo",
    title         = "{String Field Theory Solution for Any Open String Background}",
    eprint        = "1406.3021",
    archivePrefix = "arXiv",
    primaryClass  = "hep-th",
    doi           = "10.1007/JHEP10(2014)029",
    journal       = "JHEP",
    volume        = "10",
    pages         = "029",
    year          = "2014"
}

@article{Recknagel-Schomerus,
    author        = "Recknagel, Andreas and Schomerus, Volker",
    title         = "{Boundary Deformation Theory and Moduli Spaces of D-Branes}",
    eprint        = "hep-th/9811237",
    archivePrefix = "arXiv",
    primaryClass  = "hep-th",
    doi           = "10.1016/S0550-3213(99)00060-7",
    journal       = "Nucl. Phys. B",
    volume        = "545",
    pages         = "233--282",
    year          = "1999"
}

@article{erler-maccaferri-2,
    author        = "Erler, Theodore and Maccaferri, Carlo",
    title         = "{String Field Theory Solution for Any Open String Background. II}",
    eprint        = "1909.11675",
    archivePrefix = "arXiv",
    primaryClass  = "hep-th",
    doi           = "10.1007/JHEP01(2020)021",
    journal       = "JHEP",
    volume        = "01",
    pages         = "021",
    year          = "2020"
}

@article{KSM,
  author       = {Kudrna, Mat{\v{e}}j and Maccaferri, Carlo and Schnabl, Martin},
  title        = {Boundary state from {E}llwood invariants},
  journal      = {JHEP},
  year         = {2013},
  volume       = {07},
  pages        = {033},
  doi          = {10.1007/JHEP07(2013)033},
  eprint       = {1207.4785},
  archivePrefix= {arXiv},
  primaryClass = {hep-th}
}

@article{Baba-Ishibashi,
  author       = {Baba, Takayuki and Ishibashi, Nobuyuki},
  title        = {Energy from the gauge invariant observables},
  journal      = {JHEP},
  year         = {2013},
  volume       = {04},
  pages        = {050},
  doi          = {10.1007/JHEP04(2013)050},
  eprint       = {1208.6206},
  archivePrefix= {arXiv},
  primaryClass = {hep-th}
}

@article{KMPV,
  author       = {Kaja, Klaus and Maccaferri, Carlo and Portugal, Ulisses and Vo{\v{s}}mera, Jakub},
  title        = {Boundary observables in string field theory},
  year         = {2026},
  eprint       = {2607.02454},
  archivePrefix= {arXiv},
  primaryClass = {hep-th}
}

@article{MPRV,
  author       = {Maccaferri, Carlo and Poletti, Riccardo and Ruffino, Alberto and Vo{\v{s}}mera, Jakub},
  title        = {Boundary modes in string field theory},
  journal      = {JHEP},
  year         = {2025},
  volume       = {06},
  pages        = {108},
  doi          = {10.1007/JHEP06(2025)108},
  eprint       = {2502.19373},
  archivePrefix= {arXiv},
  primaryClass = {hep-th}
}

@article{MRV,
  author       = {Maccaferri, Carlo and Ruffino, Alberto and Vo{\v{s}}mera, Jakub},
  title        = {Gauge-invariant action for free string field theory with boundary},
  journal      = {JHEP},
  year         = {2026},
  volume       = {01},
  pages        = {161},
  doi          = {10.1007/JHEP01(2026)161},
  eprint       = {2506.05969},
  archivePrefix= {arXiv},
  primaryClass = {hep-th}
}

@article{StettingerBoundary,
  author       = {Stettinger, Georg},
  title        = {A boundary term for open string field theory},
  journal      = {JHEP},
  year         = {2025},
  volume       = {05},
  pages        = {226},
  doi          = {10.1007/JHEP05(2025)226},
  eprint       = {2411.15123},
  archivePrefix= {arXiv},
  primaryClass = {hep-th}
}

@article{Firat-Mamade,
  author       = {F{\i}rat, Atakan Hilmi and Mamade, Raji Ashenafi},
  title        = {Boundary terms in string field theory},
  journal      = {JHEP},
  year         = {2025},
  volume       = {02},
  pages        = {058},
  doi          = {10.1007/JHEP02(2025)058},
  eprint       = {2411.16673},
  archivePrefix= {arXiv},
  primaryClass = {hep-th}
}

@phdthesis{Kudrna,
  author       = {Kudrna, Mat{\v{e}}j},
  title        = {Level Truncation Approach to Open String Field Theory},
  school       = {Charles University, Prague},
  year         = {2019},
  eprint       = {2101.07678},
  archivePrefix= {arXiv},
  primaryClass = {hep-th}
}

@article{SchnablTachyonVacuum,
  author       = {Schnabl, Martin},
  title        = {Analytic solution for tachyon condensation in open string field theory},
  journal      = {Adv. Theor. Math. Phys.},
  year         = {2006},
  volume       = {10},
  pages        = {433--501},
  doi          = {10.4310/ATMP.2006.v10.n4.a1},
  eprint       = {hep-th/0511286},
  archivePrefix= {arXiv},
  primaryClass = {hep-th}
}

@article{GIR,
  author       = {Gaiotto, Davide and Itzhaki, Nissan and Rastelli, Leonardo},
  title        = {Closed strings as imaginary {D}-branes},
  journal      = {Nucl. Phys. B},
  year         = {2004},
  volume       = {688},
  pages        = {70--100},
  doi          = {10.1016/j.nuclphysb.2004.03.017},
  eprint       = {hep-th/0304192},
  archivePrefix= {arXiv},
  primaryClass = {hep-th}
}
\bibliographystyle{hieeetr}	
	
\end{document}